\newcommand{\beq}{\begin{equation}}
\newcommand{\eeq}{\end{equation}}
\newcommand{\bea}{\begin{eqnarray}}
\newcommand{\eea}{\end{eqnarray}}
\newcommand{\bear}{\begin{eqnarray*}}
\newcommand{\eear}{\end{eqnarray*}}

\documentclass[pra,showpacs,twocolumn,superscriptaddress,showkeys]{revtex4}

\usepackage{psfrag}
\usepackage{graphicx}

\begin{document}
\title
{EXACTLY SOLVABLE INTERACTING SPIN-ICE VERTEX MODEL}
\author{
 Anderson A. Ferreira and 
 Francisco C. Alcaraz }
\address{Instituto de F\'{\i}sica de S\~ao Carlos,\\
Universidade de S\~ao Paulo, 
CP 369, 13560-970, S\~ao Carlos, SP, Brazil}, 
\date{\today}

\begin{abstract}

A special family of solvable five-vertex model is introduced on a square
lattice.
In addition to the usual nearest neighbor interactions, the vertices defining the
model also interact along
one of the diagonals  of the lattice. Such family of models includes in
a special limit the standard six-vertex model. The exact solution of these
models gives the first application of the matrix product ansatz introduced
recently and applied successfully in the solution of quantum chains. The
phase diagram and the free energy of the models are calculated in the
thermodynamic limit. The models exhibit massless phases and our analytical
and numerical analysis indicate that such phases are governed by a conformal
field theory with central charge $c=1$ and  continuosly varying critical
exponents.
\end{abstract}

\maketitle

\section{Introduction}

        The six-vertex model was introduced by Pauling in order
 to explain the residual entropy of ice at zero temperature. Due
to its exact integrability and its connection with the XXZ quantum chain,  this
vertex model is considered as a paradigm of exact integrability 
\cite{lieb,baxter,gaudin}.
In this model the interactions are between nearest neighbor vertices
and are ruled by the geometrical connectivity of the allowed vertex
configurations on the lattice. The interaction energy among two vertices
 is zero for allowed  configurations and infinite otherwise.
In this paper  we  introduce  a special family of
five-vertex models. Besides   the usual nearest-neighbor interactions,
imposed by their connectivity, there exist additional interactions among
 vertices at larger distances. Moreover we are going to show that this 
family of models
is exactly integrable and contains as a  special case
the standard six-vertex model.

 Although the introduction of a new exactly integrable model is interesting by 
itself
we would like to mention that recently the interest on two-dimensional vertex 
models was
rised by their possible  physical realizations 
\cite{nature1}.
 Artificially produced materials have exhibited   a frustrated square-lattice 
array of
nanosize ferromagnetic domains. These materials, for large lattice spacing, 
are in an
ordered "spin-ice" square lattice where their bonds satisfy the six vertex 
connectivity rules.
By decreasing the lattice spacing, the increased frustration produces extra 
interactions among the
vertices as well as  other vertex configurations \cite{nature1}.

 The solution of the models presented on this paper are  obtained through 
the exact diagonalization
 of the diagonal-to-diagonal transfer matrix.
The exact solution of
transfer matrices associated to vertex models or quantum Hamiltonians are
usually obtained through the Bethe ansatz \cite{bethe} on its several 
formulations.
That ansatz asserts that the amplitudes of the eigenfunctions of these
operators are given by a sum of appropriate plane waves.
 Instead of making use of the  Bethe ansatz, the  solution we 
 derive will be obtained through  the matrix product
ansatz  recently introduced in  \cite{alclazo}. According to  this ansatz, 
the amplitudes
of the eigenfunctions are
given in terms of a matrix product whose matrices  obey special algebraic
relations. The present paper presents the first application of the matrix 
product
ansatz for the exact solution of a transfer matrix.

      The layout of this papers is as follows. In Sections II and III we
      introduce
the solvable family of interacting vertex models and their  associated 
diagonal-to-diagonal transfer
matrix. In Sec. IV we present the matrix product ansatz and diagonalize the 
transfer matrix obtained in the previous section.  Section V presents the 
general picture of the
roots of the spectral parameter equations associated to the exact solution
of the models. The free energy
in the thermodynamic limit is obtained in Sec. VI  whilst in Sec. VII the 
operator
content of the model on its massless phase is obtained by exploring the 
conformal
invariance of the infinite system. Finally in Sec. VIII we conclude our paper 
with a
general discussion.

\section{ The interacting five-vertex model}

\begin{figure}[h!]
\centering
{\includegraphics[angle=0,scale=0.3455 ]{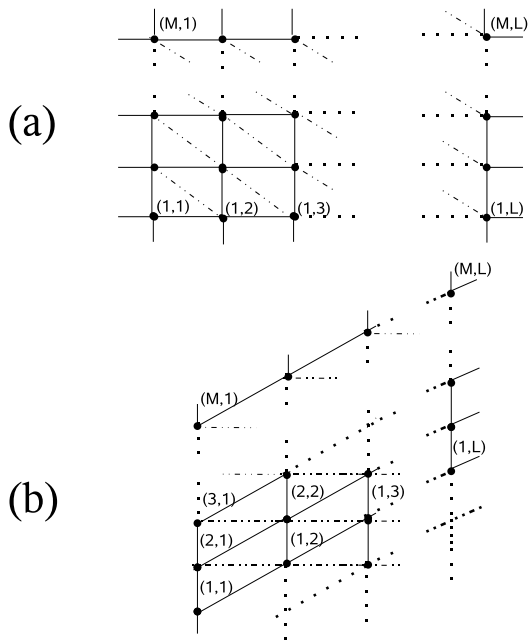}}
\caption{(a) Square lattice with $M$ horizontal ($L$ vertical) lines. The 
interacting five-vertex models have extra interactions along the dashed 
diagonals. (b) 
The deformed lattice. The dashed diagonals are now in the horizontal 
direction.
}  
\label{fig1} 
\end{figure}

\begin{figure}[h!]
\centering
{\includegraphics[angle=0,scale=0.22]{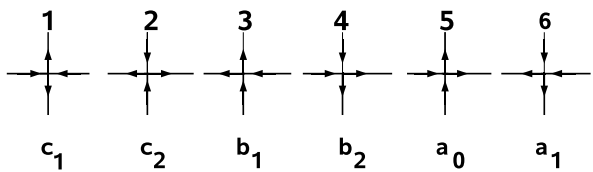}}
\caption
{The six vertex configurations allowed by the ice rules, with their respective 
fugacities.}
\label{fig2} 
\end{figure}

      The family of vertex models we  introduce and solve are
           defined on a square lattice with M rows and L columns
 (see Fig.~\ref{fig1}a). At each horizontal (vertical) link we attach an
arrow pointing to the left or right (up or down) direction. Such arrow
configurations can be equivalently described by the vertex configurations of
the lattice. A vertex configuration at a given site (center) is formed by the
four arrows attached to its links. Similar to  the six-vertex model we impose that
the allowed arrow configurations only contain vertices satisfying the ice
rules, namely, the fugacity of a given vertex is infinite unless two of its four arrows 
point inward and the two others point outward of its center.
In Fig.~\ref{fig2} we show the six vertex configurations respecting the 
ice rules, 
  with their respective fugacities
$c_1,c_2,b_1,b_2,a_0,a_1$. The partition function
 is given by the sum of all possible vertex configurations with the Boltzmann
 weights given by the product of the fugacities of the vertices. It is
 important to notice that the ice rules imply the existence of
 interactions among the nearest-neighbor vertices. These interactions
 have a zero or infinite value. For example (see Fig.~\ref{fig2}), the vertex 
with fugacity
 $a_0$ has an infinite interaction energy if the vertex on its right side is
 one of the vertices ($a_1,b_1,c_2$) or the vertex on its left side is one of
 the vertices ($a_1,b_1,c_1$). The model with   no extra interactions, 
 besides those given by the connectivity of the vertices on the lattice, is
 the well known six-vertex model. This model  is exactly integrable for arbitrary
 values of the fugacities, and is considered  a prototype  of an 
 exact integrable model \cite{lieb,baxter,gaudin}.

We consider a special family of   interacting five-vertex 
models. Besides  the previously mentioned nearest-neighbor 
interactions (infinite or zero) imposed by the lattice connectivity,
 the models also 
contain interactions among pairs of vertices at larger distances. 
The allowed vertex configurations, with their respective configurations are 
the first five configurations shown in Fig.~\ref{fig2}. Contrary to  the 
six-vertex 
model the vertex configurations with fugacity $a_1$ are  forbidden 
(zero fugacity).

Such interacting five-vertex models are labelled by a fixed positive 
integer $t$ that may take the values  $t=1,2,3,\ldots$ . This parameter   
specifies the additional interactions among 
the vertices. 
These interactions occur along the diagonals of the square lattice that 
go from the top left to the botton right direction (see the dashed diagonals 
in Fig.~\ref{fig1}a). A pair of vertices at distance $D=l\sqrt{2}$ $(l=1,2,
\ldots)$, 
in units of lattice spacing, along this diagonal interacts as follows

\noindent a) the interaction energy is zero if $l>t$,\\
\noindent b) if one of the vertices is $a_0$ the interaction energy is zero 
for all values of $l$,\\
\noindent c) if neither of the vertices is $a_0$, the interaction energy
 is infinite if $l\leq t$, except in the special case where $l=t$ and $c_2$ 
 is on the left  of $c_1$. In this case the interaction energy $e_I$ is finite 
 and produces a Boltzmann weight $c_I$ given by \cite{fn1}

\begin{eqnarray}
\label{eq:e1}
c_I=e^{-\beta e_I}=\frac{a_1}{c_1c_2}.
\end{eqnarray}

\begin{figure}[h!]
\centering
{\includegraphics[angle=0,scale=0.3028]{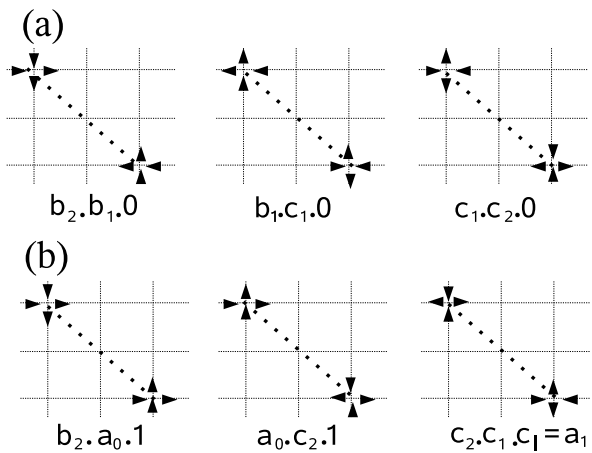}}
\caption{ Examples of forbidden (a) and allowed (b) configurations of pairs 
of vertices for the interacting vertex model with $t=2$. The contribution of 
the vertices to the Boltzmann weights are also shown. }
\label{fig3} 
\end{figure}

In Fig.~\ref{fig3}a  and Fig.~\ref{fig3}b we show, for the model with $t=2$, 
some examples 
of forbidden (infinite energy) and allowed configurations, respectively. 
 We also show, in those  figures, the 
total contribution of the pair of vertices to the Boltzmann factor. 
In general, the contribution of a given pair of vertices is zero if the pair 
is not allowed (infinite interaction energy) or is given by the product 
of their fugacities (zero interaction  energy). 
The exception to this rule happens when we have the vertices $c_1$ and $c_2$ 
at the distance $D=\sqrt{2}t$ along the diagonal, with $c_2$ on the left of
$c_1$ (see Fig.~\ref{fig3}b).  In this case, from (\ref{eq:e1}), the 
contribution is 
given by $c_1c_2c_I=a_1$.

The partition function of the vertex models are usually calculated by the 
diagonalization of the row-to-row transfer matrix, connecting the arrow 
configurations of two consecutive horizontal lines. 
In the  case under consideration, due to the existence of the extra interactions along the 
diagonals, the calculation is much simpler if we consider the 
diagonal-to-diagonal 
transfer matrix, connecting the arrow configurations of two consecutive  
diagonals of the lattice.

\section{ The diagonal-to-diagonal transfer matrix}

Following Bariev \cite{bariev} in order to construct the diagonal-to-diagonal 
transfer matrix for the interacting five-vertex models it is convenient to 
distort the square lattice shown in Fig.~\ref{fig1}a as in Fig.~\ref{fig1}b. 
In this case the 
vertices which are at closest distances along the dashed diagonals of 
Fig.~\ref{fig1}a 
are now at  the closest distance along the horizontal direction. 
We are going to solve the model with toroidal boundary conditions on the 
distorted lattice of Fig.~\ref{fig1}b. In the undeformed lattice of 
Fig.~\ref{fig1}a this 
boundary condition translates into a helical one  where we identify 
the sites

\begin{eqnarray}
\label{eq:e2}
(ij)\equiv(i+M,j)\equiv(i-L,j+L),
\end{eqnarray}

\noindent where $(i,j)$ denote the sites located at the columns $j$ $(j=1,2,
\ldots,L)$ 
and rows  $i$ ($i=1,2,\ldots, M)$.

\begin{figure}[h!]
\centering
{\includegraphics[angle=0,scale=0.29020]{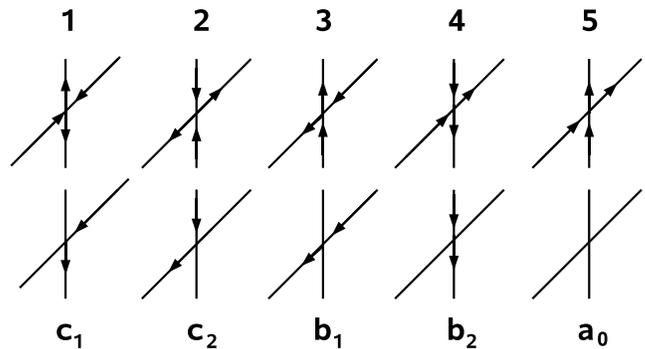}}
\caption{
The five allowed vertex configurations in the distorted diagonal lattice with 
their 
respective fugacities. In the second line we give a representation where it is 
only 
drawn  the arrows pointing down.}
\label{fig4} 
\end{figure}

In the distorted lattice of Fig.~\ref{fig1}b the allowed five vertex 
configurations 
of Fig.~\ref{fig2} are shown in Fig.~\ref{fig4}. We also represent on this 
last figure a 
convenient vertex representation where we only draw the arrows pointing to 
the down direction. Hereafter we are going to use such representation. 
We see from Fig.~\ref{fig4} that the allowed vertex configurations have 0 or 2 
arrows. 
We describe a given configuration $|\varphi>$ of $n$ arrows on a horizontal   
line of the lattice of Fig.~\ref{fig1}b by given their coordinates $(x_1,
\ldots,x_n)$ 
and a set of labels $(\alpha_1,\ldots,\alpha_n)$, where $\alpha_i=1$ or 
$\alpha_i=2$ depending if we have at the coordinates $x_i$ a vertical or 
inclined arrow respectively, i. e.,

\begin{eqnarray}
\label{eq:e3}
|\varphi>=|x_1,\alpha_1;x_2,\alpha_2;\ldots;x_n,\alpha_n>.
\end{eqnarray}

\begin{figure}[h!]
\centering
{\includegraphics[angle=0,scale=0.300]{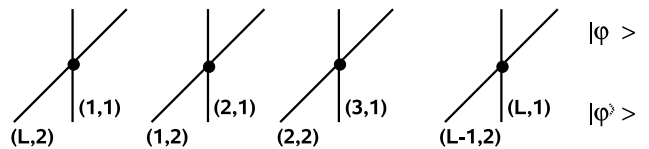}}
\caption{
Pictoric representation of the diagonal-to-diagonal transfer matrix. The 
coordinates 
($i,\alpha$) specify the possible arrows positions.}
\label{fig5} 
\end{figure}

The diagonal-to-diagonal transfer matrix $T_{D-D}$, represented pictorically 
in Fig.~\ref{fig5}, connects the arrow configurations of two horizontal lines 
in 
Fig.~\ref{fig1}b. Its matrix elements give the contribution of these 
configurations 
to the Boltzmann weight. For the allowed configurations they are  given by

\begin{eqnarray}
\label{eq:e4}
<\varphi|T_{D-D}|\varphi'>=c_1^{n_1}c_2^{n_2}
b_1^{n_3}b_2^{n_4}a_0^{n_5}c_I^{n_p},
\end{eqnarray}
\noindent where $n_i$ are the number of vertices of type $i$ $(i=1,\ldots,5)$ 
and $n_p$ is the number of pairs of vertices $c_2-c_1$ at distance of $t$ 
units along the horizontal lines (see Fig.~\ref{fig5}). 
The partition function is given by the trace

\begin{figure}[h!]
\centering
{\includegraphics[angle=0,scale=0.28]{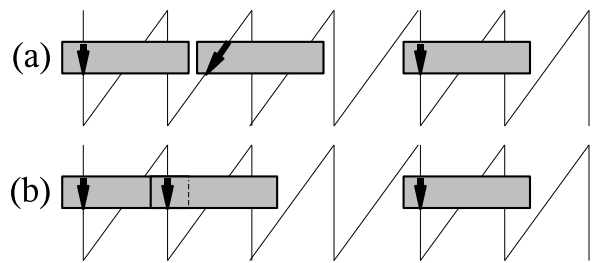}}
\caption{
Example of an allowed (a) and not allowed (b) configuration of the 
interacting five-vertex model with $t=1$. The arrows have an effective 
hard-core size $s=2t+1=3$.}
\label{fig6} 
\end{figure}

\begin{eqnarray}
\label{eq:e5}
Z=\mbox{Tr}(T_{D-D}^M).
\end{eqnarray}

It is interesting to observe that in the distorted lattice the allowed arrow
configurations on the horizontal lines, besides having a fixed number of arrows,  
could also  be interpreted as if the arrows would have an effective size 
$s=2t+1$ ($s=3,5,\ldots$), in  lattice spacing units of the distorted lattice. 
An arrow on a given link  has  hard-core interactions that exclude the 
occupation of other arrows at the link itself as well as at the $2t$-nearest 
links
on its right. In Fig.~\ref{fig6} we represent pictorically an allowed (a) and 
a not 
allowed (b) configuration of arrows for the interacting five-vertex model 
with $t=1$. This example  corresponds to arrows of effective size $s=3$.

The above interpretation, where we associate  an effective size to the arrows, 
allows us a 
simple extension of our model to the case where $t=0$. In this case the arrows 
have a unity size and the extra hard-core interactions among the vertices 
with fugacities $c_2$ and $c_1$ occur when the arrows are at the same site, 
thus giving rise to  an extra vertex with fugacity  $c_1c_2\frac{a_1}{c_1c_2}=a_1$, 
 and hence  the model reduces to the well known six-vertex model, with 
fugacities given in Fig.~\ref{fig2}.

\section{The matrix product ansatz and the diagonalization of the transfer 
matrix $T_{D-D}$}

In this section we   solve the eigenvalue equation for the 
diagonal-to-diagonal transfer matrix $T_{D-D}$ introduced in the previous 
section. As a consequence of the arrow conservation and the translation 
invariance of the arrow configurations on the horizontal lines of 
Fig.~\ref{fig1}b 
the matrix $T_{D-D}$ can be splitted into block disjoint sectors 
labelled by the number $n$ $(n=0,1,\ldots,2L)$ of arrows and momentum 
$p=\frac{2\pi}{L}j$ ($j=0,1,\ldots,L-1$). We want to solve, in each of 
these sectors, the eigenvalue equation

\begin{eqnarray}
\label{eq:e6}
\Lambda_{n,p}|\Psi_{n,p}>=T_{D-D}|\Psi_{n,p}>,
\end{eqnarray}

\noindent where $\Lambda_{n,p}$ and $|\Psi_{n,p}>$ are the eigenvalues 
and eigenvectors of $T_{D-D}$, respectively. These eigenvectors  
can be written in general as

\begin{eqnarray}
\label{eq:e7}
&&|\Psi_{n,p}>\nonumber\\
&&=\sum_{\{x\}}\sum_{\{\alpha\}}^{(*)}\phi^p_{\alpha_1,\ldots,\alpha_n}
(x_1,\ldots,x_n)|x_1,\alpha_1;\ldots;x_n,\alpha_n>,\nonumber\\
\end{eqnarray}
 where $\phi^p_{\alpha_1,\ldots,\alpha_n}(x_1,\ldots,x_n)$ is the amplitude 
 corresponding to the arrow configuration where $n$ arrows of type 
 $(\alpha_1,\ldots,\alpha_n)$ are located at $(x_1,\ldots,x_n)$,  
respectively.
 The symbol (*) in (\ref{eq:e7}) means that the 
 sums of $\{x\}$ and $\{\alpha\}$ are restricted to the sets 
 obeying the hard-core exclusions, for a given interacting parameter $t$:

\begin{equation}
\label{eq:e7.1}
x_{i+1} \geq x_i + t+1  -\delta_{\alpha_i,1}\delta_{\alpha_{i+1},2},
\end{equation}
for $i=1,\ldots,n-1$, and 
\begin{equation}
\label{eq:e7.1p} 
t+1 -\delta_{\alpha_1,1}\delta_{\alpha_n,2}
\leq x_n-x_1 \leq 
L-t-1 +\delta_{\alpha_n,1}\delta_{\alpha_1,2}.
\end{equation}
Since $|\Psi_{n,p}>$ is also an eigenvector with momentum $p$, then the amplitudes 
 satisfy

\begin{equation}
\label{eq:e8}
\frac{\phi^p_{\alpha_1,\ldots,\alpha_n}(x_1,\ldots,x_n)}{\phi^p_{\alpha_1,
\ldots,\alpha_n}(x_1+1,\ldots,x_n+1)}=
e^{-ip},
\end{equation}
for $x_n \leq L-1$, whilst for $x_n=L$
\begin{equation}
\label{eq:e8p}
\frac{\phi^p_{\alpha_1,\ldots,\alpha_n}(x_1,\ldots,L)}{\phi^p_{\alpha_1,
\ldots,\alpha_n}(1,x_1+1,\ldots,x_{n-1}+1)}=
e^{-ip}.
\end{equation}

The exact solution of ($\ref{eq:e6}$) is obtained by an appropriate ansatz 
for the unknown amplitudes $\phi^p_{\alpha_1,\ldots,\alpha_n}(x_1,\ldots,x_n)$. 
As shown in the last section our model reduces to the standard six-vertex 
model for the case where $t=0$ ($s=1$).
 An  appropriate coordinate Bethe ansatz, that solves the eigenvalue 
equation ($\ref{eq:e6}$),  is known 
in this case \cite{truong}. 
 Thus 
the 
amplitudes $\{\phi^p_{\alpha_1,\ldots,\alpha_n}(x_1,\ldots,x_n)\}$ are given, as usual,   
 by a combination of plane waves whose  wavenumbers are fixed by the 
eigenvalue equation ($\ref{eq:e6}$).

In this paper we  solve ($\ref{eq:e6}$), for general values of 
$t$ $(t=0,1,2,\ldots)$ or $s=2t+1$ ($s=1,3,5,\ldots$), by using a distinct 
ansatz. 
The matrix product ansatz 
we are going to use was introduced in \cite{alclazo} for exact integrable quantum chains. 
We present  in this paper the first application of this matrix product  ansatz 
for 
transfer matrices. According to such ansatz the amplitudes 
$\phi^p_{\alpha_1,\ldots,\alpha_n}(x_1,\ldots,x_n)$ are given in terms 
of a matrix product   satisfying an unknown associative algebra. 
The model is exactly integrable if the 
 eigenvalue equations fix consistently  the algebraic relations among the 
matrices.

In order to formulate the matrix product ansatz we make a one-to-one 
correspondence between configurations of arrows and  products of matrices. 
The matrix product associated to a given arrow configuration 
is obtained by associating a matrix $E$ to the sites with no arrow,  a 
 matrix $A^{(\alpha)}$  to the sites with a single arrow of type 
$\alpha$ ($\alpha=1,2$), and finally 
  the matrix $A^{(1)}E^{-1}A^{(2)}$ to the sites with two arrows \cite{fn2}.
 The unknown amplitudes in (\ref{eq:e7}) are obtained by  
 associating them to the matrix product ansatz 

\begin{eqnarray}
\label{eq:e9}
&&\phi^p_{\alpha_1,\ldots,\alpha_n}(x_1,\ldots,x_n) \Leftrightarrow
E^{x_1-1}A^{(\alpha_1)}E^{x_2-x_1-1}A^{(\alpha_2)}\nonumber\\&&\cdots 
E^{x_n-x_{n-1}-1}A^{(\alpha_n)}E^{L-x_n}.
\end{eqnarray}

Actually $E$ and $A$ are abstract operators with an associative product. 
A well defined eigenfunction is obtained, apart from a normalization factor, 
if all the amplitudes are  uniquely related. Equivalently, in the subset 
of words (products of matrices) 
of the algebra containing $n$ matrices $A$ and $(L-n)$ 
matrices $E$, there exists only a single independent word. The relation 
between any two words gives the ratio between the corresponding amplitudes 
in (\ref{eq:e9}). 

We could  also formulate the ansatz (\ref{eq:e9}) by associating a complex 
number to the single independent word. We can choose any operation for the 
matrix products that gives a non-zero scalar. In  the original formulation of 
the matrix product ansatz for exactly integrable quantum chains 
with periodic boundary conditions \cite{alclazo},
  the trace operation was chosen to produce the following  scalar

\begin{eqnarray}
\label{eq:e9p}
&&\phi^p_{\alpha_1,\ldots,\alpha_n}(x_1,\ldots,x_n) =
\mbox{Tr}
[E^{x_1-1}A^{(\alpha_1)}E^{x_2-x_1-1}A^{(\alpha_2)}\nonumber\\&&\cdots 
E^{x_n-x_{n-1}-1}A^{(\alpha_n)}E^{L-x_n}\Omega_p].
\end{eqnarray}
The matrix $\Omega_p$ was chosen to have a given algebraic 
relation with the matrices $E$ and $A$.  Recently, Golinelli and 
Mallick \cite{golimalli} have show that in the particular case of the 
asymmetric exclusion problem in a periodic chain it is possible to 
formulate the ansatz only by imposing relations between the matrices 
$E$ and $\Omega_p$. The relations between $A$ and $\Omega_p$ is then totally 
arbitrary. Actually, as we are going to show in this paper, we do not need 
to impose any algebraic relation between the matrices $E$ and $A$ with 
$\Omega_p$. The matrix $\Omega_p$ can be   any arbitrary matrix 
that produces 
a non vanishing trace  in (\ref{eq:e9p}). This observation is not particular 
to the present model. It is valid for any of the exactly integrable quantum 
chains solved in the original formulation of the matrix product 
ansatz \cite{alclazo}. Instead of restricting our matrix product 
ansatz with the 
trace operation, as in (\ref{eq:e9p}), 
we will consider the more general formulation  given by (\ref{eq:e9}). 

Since the eigenfunctions produced by the ansatz have a well defined momentum 
$p = \frac{2\pi}{L}j$ ($j=0,\ldots,L-1$), the relations (\ref{eq:e8}) and 
(\ref{eq:e8p}) imply the following constraints  for the matrix products 
appearing in the ansatz (\ref{eq:e9})
\begin{eqnarray}
\label{m1}
&&E^{x_1-1}A^{(\alpha_1)}E^{x_2 -x_1 -1} \cdots A^{(\alpha_n)} 
E^{L-x_n}  =  \nonumber \\ 
&&  e^{-ip} E^{x_1}A^{(\alpha_1)}E^{x_2-x_1-1}\cdots A^{(\alpha_n)} 
E^{L-x_n-1},
\end{eqnarray}
for $x_n \leq L-1$, and for $x_n =L$ 
\begin{eqnarray}
\label{m2}
&&E^{x_1-1}A^{(\alpha_1)}E^{x_2 -x_1 -1} \cdots A^{(\alpha_n)} 
  =  \nonumber \\ 
&&  e^{-ip}A^{(\alpha_n)}E^{x_1-1}A^{(\alpha_1)}\cdots A^{(\alpha_{n-1})} 
E^{L-x_{n-1}-1}.
\end{eqnarray}

In order to proceed, in the usual way, we are going to consider firstly  the 
eigensectors of $T_{D-D}$ 
with small values of $n$. 

\noindent {$n=0$}. We have the trivial solution presenting  a single
eigenvalue $\Lambda_0=a_0^L$.

\noindent{$n =1$}. We have in this case $2L$ possible configurations, 
corresponding to a single arrow of type $\alpha=1,2$ at  any lattice 
position $x=1,\ldots,L$. The eigenvalue equation ($\ref{eq:e6}$) 
gives us the $2L$ equations

\begin{eqnarray}
\label{eq:e11}
&&\frac{\Lambda _1}{a_0^{L-1}}E^{x-1}A^{(1)}E^{L-x}=
\nonumber \\
b_2 
  E^{x-1}A^{(1)} 
&&  E^{L-x}
+c_1E^{x-1}A^{(2)}E^{L-x},
\end{eqnarray}
for $x =1,\ldots,L$,

\begin{eqnarray}
\label{eq:e12}
&&\frac{\Lambda _1}{a_0^{L-1}}E^{x-1}A^{(2)}E^{L-x}=
\nonumber \\
c_2E^xA^{(1)} 
&&  E^{L-x-1}
+b_1E^xA^{(2)}E^{L-x-1},
\end{eqnarray}
for $x=1,\ldots,L-1$, and for $x=L$
\begin{eqnarray}
\label{eq:e12p}
&&\frac{\Lambda _1}{a_0^{L-1}}E^{L-1}A^{(2)}=
c_2A^{(1)}E^{L-1}  \nonumber \\ 
&& +
b_1A^{(2)}E^{L-1}.
\end{eqnarray}
  These equations are written in a convenient form by 
 imposing that both matrices $A^{\alpha}$ ($\alpha=1,2$) depend on a single 
 spectral 
 parameter matrix $A_k$:

\begin{eqnarray}
\label{eq:e13}
A^{(1)}=\phi_1^1A_kE^{1-2t},\;\;\;\;A^{(2)}=\phi_2^1A_kE^{1-2t},
\end{eqnarray}
 where $\phi_1^1$ and $\phi_2^1$ are unknown scalars and $A_k$ satisfies  the 
following algebraic relation with the matrix $E$

\begin{eqnarray}
\label{eq:e14}
EA_k=e^{ik}A_kE.
\end{eqnarray}

Inserting ($\ref{eq:e13}$) on ($\ref{eq:e11}$) we obtain

\begin{eqnarray}
\label{eq:e15}
&&(\Lambda_1\phi_1^1-b_2a_0^{L-1}\phi_1^1-c_1a_0^{L-1}\phi_2^1)\nonumber\\
&&\times E^{x-1}A_kE^{1-2t}E^{L-x}=0.
\end{eqnarray}

On the other hand inserting 
($\ref{eq:e13}$) on ($\ref{eq:e12}$)-(\ref{eq:e12p})  
and using (\ref{eq:e14}) we obtain

\begin{eqnarray}
\label{eq:e16}
&&(\Lambda_1\phi_2^1-c_2a_0^{L-1}\phi_1^1e^{ik}-b_1a_0^{L-1}\phi_2^1e^{ik})\nonumber\\
&&\times E^{x-1}A_kE^{1-2t}
E^{L-x}=0,
\end{eqnarray}
for $x=1,\ldots,L-1$ and

\begin{eqnarray}
\label{18p}
&&(\Lambda_1e^{ikL}\phi_2^1-c_2a_0^{L-1}\phi_1^1e^{ik}-b_1a_0^{L-1}\phi_2^1e^{ik})\nonumber\\
&&\times A_kE^{L-2t}=0.
\end{eqnarray}
However by inserting (\ref{eq:e14}) in (\ref{m1}) we verify that $k$ coincide 
with the momentum of the eigenstate:
\begin{equation}
\label{18pp}
k = p = \frac{2\pi m}{L}, \; \; \; m=0,1,\ldots,L-1.
\end{equation}
In order to produce  a non-zero norm state we should impose 
$E^{x-1}A_kE^{1-2t}E^{L-x}\neq 0$, for 
$x=1,\ldots,L$.  Due to  the equality $e^{ikL}=1$,  
equations (\ref{eq:e15})-(\ref{18p})
 reduce to the eigenvalue 
problem of a $2\times 2$ matrix

\beq
\label{eq:e19}
\frac{\Lambda _1}{a_0^{L-1}}
\left[\matrix{ \phi_1^1 \cr \phi_2^1}\right] = 
\left[ \matrix{ b_2 & c_1 \cr c_2e^{ik} & b_1e^{ik}}\right] 
\left[\matrix{ \phi_1^1 \cr \phi_2^1}\right] . 
\eeq

The diagonalization of (\ref{eq:e19}) gives us  the 
eigenvalue

\begin{eqnarray}
\label{eq:e20}
&&\Lambda_1(k)= \Lambda _1^{(l)}(k)= 
\frac{a_0^{L-1}}{2}(b_2+b_1e^{ik} \nonumber \\
&&+l[(b_2+b_1e^{ik})^2-4e^{ik}
(b_2b_1-c_2c_1)]^{\frac{1}{2}}),
\end{eqnarray}
 with $l=\pm 1$ and $k$ given by (\ref{18pp}). 

\noindent {$n=2$}. The eigenvalue equation ($\ref{eq:e6}$) gives  two types 
of relations 
for the amplitudes ($\ref{eq:e9}$). The configuration where we have no 
collisions, i. e, $|x_1,\alpha_1;x_2,\alpha_2> \neq |x_1,2;x_1+t+1,1>$ gives 
generalizations of the equations ($\ref{eq:e11}$)-($\ref{eq:e12p}$) 
for two particles

\begin{eqnarray}
\label{23.1}
&&\frac{\Lambda _2}{a_0^{L-2}}E^{x_1-1}A^{(1)}E^{x_2-x_1-1}A^{(1)}E^{L-x_2}
=\nonumber\\
&&b_2^2E^{x_1-1}A^{(1)}E^{x_2-x_1-1}A^{(1)}E^{L-x_2}
\nonumber\\
&&+c_1b_2E^{x_1-1}A^{(2)}E^{x_2-x_1-1}A^{(1)}E^{L-x_2}
\nonumber\\
&&+b_2c_1E^{x_1-1}A^{(1)}E^{x_2-x_1-1}A^{(2)}E^{L-x_2}
\nonumber\\
&&+c_1^2E^{x_1-1}A^{(2)}E^{x_2-x_1-1}A^{(2)}E^{L-x_2}
,
\end{eqnarray}

\begin{eqnarray}
\label{23.2}
&&\frac{\Lambda _2}{a_0^{L-2}}E^{x_1-1}A^{(2)}E^{x_2-x_1-1}A^{(1)}E^{L-x_2}
=\nonumber\\
&&c_2b_2E^{x_1}A^{(1)}E^{x_2-x_1-2}A^{(1)}E^{L-x_2}
\nonumber\\
&&+b_1b_2E^{x_1}A^{(2)}E^{x_2-x_1-2}A^{(1)}E^{L-x_2}
\nonumber \\
&&+c_2c_1[E^{x_1}A^{(1)}E^{x_2-x_1-2}A^{(2)}E^{L-x_2}
\nonumber\\
&&+b_1c_1E^{x_1}A^{(2)}E^{x_2-x_1-2}A^{(2)}E^{L-x_2}
,
\end{eqnarray}

\begin{eqnarray}
\label{23.3}
&&\frac{\Lambda _2}{a_0^{L-2}}E^{x_1-1}A^{(1)}E^{x_2-x_1-1}A^{(2)}E^{L-x_2}=
\nonumber\\
&&c_2b_2E^{x_1-1}A^{(1)}E^{x_2-x_1}A^{(1)}E^{L-x_2-1}
\nonumber\\
&&+c_1c_2[E^{x_1-1}A^{(2)}E^{x_2-x_1}A^{(1)}E^{L-x_2-1}
\nonumber\\
&&+b_2b_1E^{x_1-1}A^{(1)}E^{x_2-x_1}A^{(2)}E^{L-x_2-1}
\nonumber\\
&&+c_1b_1E^{x_1-1}A^{(2)}E^{x_2-x_1}A^{(2)}E^{L-x_2-1}
,
\end{eqnarray}

\begin{eqnarray}
\label{23.4}
&&\frac{\Lambda _2}{a_0^{L-2}}E^{x_1-1}A^{(2)}E^{x_2-x_1-1}A^{(2)}E^{L-x_2}
=\nonumber\\
&&c_2^2E^{x_1}A^{(1)}E^{x_2-x_1-1}A^{(1)}E^{L-x_2-1}
\nonumber\\
&&+b_1c_2E^{x_1}A^{(2)}E^{x_2-x_1-1}A^{(1)}E^{L-x_2-1}
\nonumber \\&&+c_2b_1E^{x_1}A^{(1)}E^{x_2-x_1-1}
A^{(2)}E^{L-x_2-1}\nonumber\\
&&+b_1^2E^{x_1}A^{(2)}E^{x_2-x_1-1}A^{(2)}E^{L-x_2-1}.
\end{eqnarray}
In the case where $x_2=L$, equations (\ref{23.3}) and (\ref{23.4}) are 
replaced by 

\begin{eqnarray}
\label{23.5}
&&\frac{\Lambda _2}{a_0^{L-2}}E^{x_1-1}A^{(1)}E^{L-x_1-1}A^{(2)}
=\nonumber\\
&&c_2b_2A^{(1)}E^{x_1-2}A^{(1)}E^{L-x_1}
\nonumber\\
&&+c_1c_2A^{(1)}E^{x_1-2}A^{(2)}E^{L-x_1}
\nonumber\\
&&+b_2b_1A^{(2)}E^{x_1-2}A^{(1)}E^{L-x_1}
\nonumber\\
&&+c_1b_1A^{(2)}E^{x_1-2}A^{(2)}E^{L-x_1}
,
\end{eqnarray}

\begin{eqnarray}
\label{23.6}
&&\frac{\Lambda _2}{a_0^{L-2}}E^{x_1-1}A^{(2)}E^{L-x_1-1}A^{(2)}
=\nonumber\\
&&c_2^2A^{(1)}E^{x_1-1}A^{(1)}E^{L-x_1-1}
\nonumber\\
&&+b_1c_2A^{(1)}E^{x_1-1}A^{(2)}E^{L-x_1-1}
\nonumber\\
&&+c_2b_1A^{(2)}E^{x_1-1}A^{(1)}E^{L-x_1-1}
\nonumber\\
&&+b_1^2A^{(2)}E^{x_1-1}A^{(2)}E^{L-x_1-1}
.
\end{eqnarray}

On the other hand, the configurations $|x_1,2;x_2=x_1+t+1,1>$, where we have 
 collision of the arrows, give us, instead,  the  equations

\begin{eqnarray}
\label{eq:e24}
&&\frac{\Lambda_2}{a_0^{L-2}}E^{x_1-1}A^{(2)}E^{t}A^{(1)}E^{L-x_1-t-1}
=\nonumber\\
&&a_1E^{x_1}A^{(1)}E^{t-1 }A^{(2)}E^{L-x_1-t-1},
\end{eqnarray}

\noindent where we have used the interaction energy  $e_I=\frac{a_1}{c_1c_2}$, 
given by (\ref{eq:e1}).
As in the case $n=1$ we identify the matrices $A^{\alpha}$ ($\alpha=1,2)$ as 
composed by two spectral dependent matrices $A_{k_1}$ and $A_{k_2}$:

\begin{eqnarray}
\label{eq:e25}
A^{(\alpha)}=\sum_{i=1}^2\phi_{\alpha}^iA_{k_i}E^{1-2t},
\end{eqnarray}
 where  $\phi_{\alpha}^i$ $(\alpha ,i=1,2)$ are unknown scalars
 and the spectral dependent matrices obey the commutation relation with the 
matrix $E$

\begin{eqnarray}
\label{eq:e26}
EA_{k_j}=e^{ik_j}A_{k_j}E, \;\;\; j=1,2.
\end{eqnarray}
The relations (\ref{eq:e25}) and (\ref{eq:e26}),  when inserted  in 
 (\ref{m1}) with $n=2$, relate the momentum of the 
eigenstate with the spectral parameters:
\begin{equation}
\label{26p}
p = k_1 +k_2.
\end{equation}
Using (\ref{eq:e25})-(\ref{26p}) in 
(\ref{23.1})-({\ref{23.6}) the non-zero norm condition  for the 
eigenstate gives, for any $x_1$ and $x_2$, a single set of equations for 
the scalars 
 $\phi_{\alpha}^i$ $(\alpha ,i=1,2)$:

\small

\bea
\label{eq:27}
&&\frac{\Lambda_2}{a_0^{L-2}}
\left[\matrix{
\phi_ 1 ^1\phi_ 1 ^2\cr
\phi_ 1 ^1\phi_ 2 ^2\cr
\phi_ 2 ^1\phi_ 1 ^2\cr
\phi_ 2 ^1\phi_ 2 ^2\cr}
\right]=\nonumber\\
&&\left[\matrix{
b_2^2  &b_2c_1 &c_1b_2 &c_1^2\cr
b_2c_2e^{ik_2}  &b_2b_1e^{ik_2} &c_1c_2e^{ik_2} &c_1b_1e^{ik_2}\cr
c_2b_2e^{ik_1}  &c_2c_1e^{ik_1} &b_1b_2e^{ik_1} &b_1c_1e^{ik_1}\cr
c_2^2e^{i(k_1+k_2)}  &c_2b_1e^{i(k_1+k_2)}& b_1c_2e^{i(k_1+k_2)}& 
b_1^2e^{i(k_1+k_2)}\cr}
\right]\nonumber\\
&&\times\left[\matrix{
\phi_ 1 ^1\phi_ 1 ^2\cr
\phi_ 1 ^1\phi_ 2 ^2\cr
\phi_ 2 ^1\phi_ 1 ^2\cr
\phi_ 2 ^1\phi_ 2 ^2\cr}
\right].
\eea

\normalsize

This last equation can be rewritten in the tensorial form as

\small

\begin{eqnarray}
\label{eq:28}
&&\frac{\Lambda_2}{a_0^{L-2}}
\left( \left[\matrix{
\phi_ 1 ^1\cr
\phi_ 2 ^1\cr}
\right]
\otimes 
\left[\matrix{
\phi_ 1 ^2\cr
\phi_ 2 ^2\cr}
\right]\right) =\nonumber\\
&&\left(\left[\matrix{
b_2 &c_1 \cr
c_2e^{ik_1} &b_1e^{ik_1} \cr}
\right]
\otimes   
\left[\matrix{
b_2 &c_1 \cr
c_2e^{ik_2} &b_1e^{ik_2} \cr}
\right]\right)
\left( \left[\matrix{
\phi_ 1 ^1\cr
\phi_ 2 ^1\cr}
\right]
\otimes 
\left[\matrix{
\phi_ 1 ^2\cr
\phi_ 2 ^2\cr}
\right]\right).\nonumber\\
\end{eqnarray}

\normalsize
We can always choose one of the fugacities to be  unity. 
If we set $a_0=1$ we see from ($\ref{eq:e19}$) that the eigenvalue 
$\Lambda_2(k_1,k_2)$ is given by

\begin{eqnarray}
\label{eq:e29}
\Lambda_2(k_1,k_2)=\Lambda_1(k_1)\Lambda_1(k_2),
\end{eqnarray}
 where $\Lambda_1(k)$ is the one-particle energy ($\ref{eq:e20}$), and the 
scalars $\phi^i_{\alpha}$ satisfy

\begin{eqnarray}
\label{eq:e30}
\phi^i_1=\frac{c_1\phi^i_2}{\Lambda_1(k_i)-b_2},\;\;\; i=1,2.
\end{eqnarray}

 Up to now the commutations relations of the matrices $A_{k_i}$ among 
themselves are not known. They are going to be fixed by the "colliding" 
equations ($\ref{eq:e24}$). Inserting ($\ref{eq:e25}$) in ($\ref{eq:e24}$) 
and using ($\ref{eq:e26}$) we obtain
\begin{eqnarray}
\label{eq:e32}
&&\sum_{i,j=1,i\neq j}^2(\Lambda_2\phi^i_2\phi^j_1e^{ik_j}-a_1\phi^i_1\phi^j_2
e^{i(k_i+k_j)})\nonumber\\
&&E^{x_1-1}A_{k_i}E^{-t}A_{k_j}E^{L-x_1-3t+1}=0,
\end{eqnarray}
 where $\Lambda_2$ is given by ($\ref{eq:e29}$). This  last equation 
 can be rewritten conveniently by the use of  the relations
\small

\begin{eqnarray}
\label{eq:e33}
-\Lambda_1^2(k_j)+b_2\Lambda_1(k_j)=
-\Lambda_1(k_j)b_1e^{ik_j}+e^{ik_j}(b_2b_1-c_2c_1)
\end{eqnarray}
\normalsize

\noindent with $j=1,2,$ and

\begin{eqnarray}
\label{eq:e34}
\frac{\phi_2^1\phi_1^2}{\phi_2^2\phi_1^1}=\frac{b_2-\Lambda_1(k_1)}
{b_2-\Lambda_1(k_2)},\end{eqnarray}
 that follows from ($\ref{eq:e19}$) and ($\ref{eq:e30}$), 
respectively. These last two relations imply the commutation relations among 
the spectral matrices

\begin{eqnarray}
\label{eq:e35}
A_{k_j}A_{k_i}=s(k_j,k_i)A_{k_i}A_{k_j},\;\;\; A_{k_i}^2=0,\;\;(i=1,2)
\end{eqnarray}
 where

\begin{eqnarray}
\label{eq:e36}
&&s(k_j,k_i)=\nonumber\\
&&-\frac{\Lambda_1(k_i)\Lambda_1(k_j)b_1-
\Lambda_1(k_j)(b_2b_1-c_2c_1+a_1)+a_1b_2}
{\Lambda_1(k_i)\Lambda_1(k_j)b_1-\Lambda_1(k_i)(b_2b_1-c_2c_1+a_1)+a_1b_2}.
\nonumber\\
\end{eqnarray}

The spectral parameters $k_1$ and $k_2$, which were free up to now, 
are going to be fixed by the boundary relation (\ref{m2}). 
Inserting (\ref{eq:e25})  
in (\ref{m2}) and using the commutation relations (\ref{eq:e26}) we obtain
\begin{eqnarray}
\label{27.1}
&&\sum_{j,l=1}^2 \phi_{\alpha_1}^j\phi_{\alpha_2}^l e^{ik_j(x_1-1)} 
e^{k_l(L-2t-1)} A_{k_j}A_{k_l}E^{L-4t} =  \nonumber \\
&&e^{-ip}\sum_{j,l=1}^2 \phi_{\alpha_2}^j\phi_{\alpha_1}^l e^{ik_l(x_1-2t)} 
 A_{k_j}A_{k_l}E^{L-4t}. 
\end{eqnarray}
Making the permutation $j\leftrightarrow l$ on the right-hand side  of the 
last equation and 
using (\ref{eq:e35}) we obtain the spectral parameter equations that fix 
$k_1$ and $k_2$:

\begin{eqnarray}
\label{eq:e37}
e^{ik_jL}=\left(\frac{e^{ik_j}}{e^{ik_l}}\right)^{2t}s(k_j,k_l),\;\;\; j\neq 
l=1,2.
\end{eqnarray}
The solutions $(k_1,k_2)$ of these equations when inserted in 
($\ref{eq:e29}$) give us the eigenvalues $\Lambda_2(k_1,k_2)$.

$n>2$. The general case follows straightforwardly from the $n=2$ case. 
The equations  with no collisions lead us to identify the matrices 
$A^{(\alpha)}$ of the ansatz ($\ref{eq:e9}$) as being composed by $n$ spectral 
dependent matrices $A_{k_j}$ $(j=1,\ldots,n)$,

\begin{eqnarray}
\label{eq:e38}
A^{\alpha}=\sum_{j=1}^n\phi_{\alpha}^jA_{k_j}E^{1-2t},\;\;\; \alpha=1,2,
\end{eqnarray}
 satisfying the algebraic relation

\begin{eqnarray}
\label{eq:e39}
EA_{k_j}=e^{ik_j}A_{k_j}E.
\end{eqnarray}
The equation (\ref{m1}) relates  the momentum $p$ of the 
eigenstate and the spectral parameters:
\begin{equation}
\label{40p}
p = \sum_{j=1}^n k_j.
\end{equation}
The equations comming from the configurations 
where two arrows are at colliding positions fix the 
commutations relations among the matrices $A_{k_j}$:

\begin{eqnarray}
\label{eq:e40}
A_{k_j}A_{k_l}=s(k_j,k_l)A_{k_l}A_{k_j},\;\;\; A_{k_j}^2=0,\;\;\;j=1,\ldots,n,
\end{eqnarray}
 where $s(k_j,k_l)$ is given by ($\ref{eq:e36}$). 
 The equations with more than two arrows  at collisions give 
 relations that are automatically satisfied, due to the 
 associativity of the algebra of the matrices $A_{k_j}$ $(j=1,\ldots,n)$. 
 The eigenvalues of the transfer matrix are given by

\begin{equation}
\label{eq:e41}
\Lambda_n(k_1,\ldots,k_n)=\prod_{j=1}^n\Lambda_1(k_j).
\end{equation}
  The spectral parameters $\{k_i\}$ are fixed by the boundary relation 
(\ref{m2}) 
 and are given by  the solution of the nonlinear set of equations

\begin{eqnarray}
\label{eq:e42}
e^{ik_jL}=-\prod_{l=1}^ns(k_j,k_l)
\left(\frac{e^{ik_j}}{e^{ik_l}}\right)^{2t},\;\;\;j=1,\ldots,n,
\end{eqnarray}
 with $s(k_j,k_l)$ given by ($\ref{eq:e36}$). This last equation 
 reproduces for $t=0$ the spectral parameter equations for the six-vertex 
 model obtained in \cite{truong} by using  the coordinate Bethe ansatz.

 Since $p=\sum_{j=1}^nk_j$ we can rewrite the spectral parameter equations 
 ($\ref{eq:e42}$),  
  in the sector with a number $n$ ($n=1,2,\ldots)$ 
 of arrows and momentum $p=\frac{2\pi j}{L}$ ($j=0,1,\ldots,L-1)$,  as 

\begin{eqnarray}
\label{eq:e43}
e^{ik_j(L-2tn)}e^{-2ipt}=-\prod_{l=1}^ns(k_j,k_l).
\end{eqnarray}
 Consequently the eigenvalues belonging to  the sector labelled by 
 ($n,p$) of $T_{D-D}$ of the interacting five-vertex model with a 
 parameter $t$   ($t=0,1,2,\ldots$)
 are  related to those of the standard six-vertex 
 model $(t=0)$. The six-vertex model is defined on a lattice  size $L^{'}=L-2nt$ and with a 
 momentum dependent seam \cite{fn3} 
 along the vertical direction of Fig.~\ref{fig1}b.  
  The same phenomenon  also happens on quantum Hamiltonians 
  with hard-core exclusion effects \cite{alcbar}.

We should mention  that is also possible to
reobtain our solution of the interacting vertex models by using the
coordinate Bethe ansatz. The  eigenfunctions we
derived through the matrix product ansatz give the educated guess that we
should use for the   coordinate Bethe ansatz. The spectral parameter 
equations 
we  obtain are the same as (\ref{eq:e42}).

Before closing this section let us give a possible representation 
for the matrices 
$A$ and $E$ of the ansatz (\ref{eq:e9}). For a given solution 
$\{k_1,\ldots,k_n\}$ of the spectral parameter equations (\ref{eq:e43}), in 
the 
sector with $n$ particles, the matrices $E$ and $\{A_{k_1},\ldots,A_{k_n}\}$ 
have 
the following finite-dimensional representation
\begin{eqnarray}
\label{repre}
&&E = 
\bigotimes_{l=1}^n  \left(\matrix{1 & 0 \cr 0 &e^{-ikl}}\right), \nonumber \\
&&A_{k_j} = \left[\bigotimes_{l=1}^{j-1}\left(
\matrix{s(k_j,k_l) &0 \cr 0& 1}\right) \right]
\otimes \left( \matrix{ 0 & 1 \cr 0 & 0}\right) \nonumber \\
&& \bigotimes_{l=j+1}^n \left( \matrix{ 1 & 0 \cr 0 & 1 } \right),
\end{eqnarray}
where $s(k_j,k_l)$ are given by (\ref{eq:e36}) and $A$ is obtained from  
(\ref{eq:e38}).  The dimension of the representation is $2^n$ and the products 
appearing on the ansatz are traceless. If we want a 
formulation of the  matrix product ansatz where the trace operation is used, 
as in the formulation (\ref{eq:e9p}), it is quite simple to produce the 
matrix $\Omega_p$ that gives a non-zero value for the trace. We should stress 
that  in the original formulation of the ansatz \cite{alclazo}, it was 
required unnecessary 
additional algebraic relations of the matrices $E$ and $A$ with the 
boundary matrix $\Omega_p$. The representations, in this case, are probably 
 infinite dimensional. 

\section{Roots of the spectral parameter equations}

In order to complete the solution of any integrable model we need to 
find the roots of the associated spectral parameter equations 
(Eq. ($\ref{eq:e42}$) in our case). The solution of those equations is in 
general a quite difficult problem for finite $L$. However  numerical analysis 
on small lattices allows us to conjecture,   
 for each problem,  
the particular distributions of roots that correspond to  the most 
important eigenvalues in the bulk limit ($L \rightarrow \infty$).
Those are the eigenvalues with larger real part  in 
the case of transfer matrix calculations. 
The equations we obtained in the last section were never analyzed 
previously either for finite or infinite values of $L$. 
Even in the simplest 
case $t=0$, where the model reduces to the six-vertex model, 
the spectral parameter equations obtained in \cite{truong}, through the Bethe 
ansatz,  were not analyzed in the literature.
A detailed numerical analysis of the Bethe ansatz equations for the row-to-row
transfer matrix of the six-vertex model, or the XXZ chain was previously done 
\cite{ab2,
batchelor}. However, as we saw in last section, the corresponding equations 
for
the
diagonal-to-diagonal transfer matrix of the six-vertex model are quite 
different
 and
a  numerical study of these equations is still missing.

In our general solution of last section we have, for arbitrary 
values of the interacting range $t$ ($t=0,1,2,\ldots)$, 
five free parameters: 
$(a_1,b_1,b_2,c_1,c_2)$. The particular case where we have no interactions 
along the diagonals ($a_1 =0$) is special and is not going to be considered 
here 
(see \cite{wu} for a dicussion of the parameterization in this case).

In order to simplify our analysis we are going 
hereafter to restrict ourselves to a symmetric version of our model 
with only three free parameters ($\delta,b,c$), namely,

\begin{eqnarray}
\label{eq:r1}
a_0=1,\;\;\;a_1=\delta^2,\;\;\;b_1=b_2=b\delta,\;\;\;c_1=c_2=c\delta .
\end{eqnarray}
The parameter $\delta$ gives, in the case where $t=0$, the contribution 
to the fugacity due to an electric field on the symmetric six-vertex model. 
 For general values of $t$, $\mu =-\ln \delta$,  plays the role of a 
chemical potential controlling the number of arrows in the thermodynamic 
limit.

Instead of writing the spectral parameter 
equations in terms of the spectral parameters ($k_1,\ldots,k_n)$ as 
in ($\ref{eq:e43}$), it is more convenient to write these equations in 
terms of the variables $\lambda_j\equiv \frac{\Lambda_1(k_j)}{\delta}$, 
with $\Lambda_1(k_j)$ given by ($\ref{eq:e20}$). In this case the 
eigenvalues of $T_{D-D}$ are given by

\begin{eqnarray}
\label{eq:r2}
\Lambda_n=\delta^n\lambda_1\cdots\lambda_n,
\end{eqnarray}
 where $\{\lambda_j\}$ satisfy

\begin{eqnarray}
\label{eq:r3}
&&\left(\frac{\lambda_j(b-\lambda_j)}{b(b-\lambda_j)-c^2}\right)
^{L-2tn}e^{-2ipt}=\nonumber\\
&&(-)^{n+1}\prod_{l=1}^n\frac{\lambda_l\lambda_j-2\Delta\lambda_j+1}
{\lambda_l\lambda_j-2\Delta\lambda_l+1},\;\;\; j=1,\ldots,n,
\end{eqnarray}
 and  we have introduced the anisotropy parameter

\begin{eqnarray}
\label{eq:r4}
\Delta=\frac{b^2-c^2+1}{2b}.
\end{eqnarray}

We see, from ($\ref{eq:r2}$)-($\ref{eq:r4}$), that we have now only 
two free parameters: $b$ and $c$. The interacting parameter $\delta$,  
that gives the contribution due to the interactions  among the vertices 
along the diagonal, does not appear on the equations ($\ref{eq:r3}$) 
and  ($\ref{eq:r4}$). It only gives an overall scale for the 
eigenvalues as shown in ($\ref{eq:r2}$). Inspired on the usual 
parameterization of the six-vertex model \cite{baxter} we express the 
parameters $b$ and $c$ in terms of the parameters $\sigma$ and $\gamma$: 

\begin{eqnarray}
\label{eq:r5}
&&b=b(\gamma,\sigma)=\frac{\sin\sigma}{\sin(\gamma-\sigma)},\;\;\;
c=c(\gamma,\sigma)=\frac{\sin\gamma}{\sin(\gamma-\sigma)},\nonumber\\
&&\Delta=-\cos \gamma.
\end{eqnarray}

The fact that  $\Delta$ is a real number implies that $\gamma$ is real for
$-1\leq \Delta \leq 1$, 
and  pure imaginary  for $|\Delta |> 1$. Since the right-hand 
side of ($\ref{eq:r3}$) is the same for all values of the parameter $t$
it is 
interesting, as in the six-vertex model \cite{truong}, to make the change of 
variables $\lambda_j\rightarrow \sigma_j$, where,

\begin{eqnarray}
\label{eq:r6}
\lambda_j=\frac{\sinh(i\gamma-\sigma_j)}{\sinh\sigma_j},\;\;\; j=1,\ldots,n.
\end{eqnarray}
In terms of these new variables $\{\sigma_j\}$ the spectral parameter 
equations ($\ref{eq:r3}$) become

\begin{eqnarray}
\label{eq:r7}
&&\left(\frac{\sinh(i\gamma-\sigma_j)\sinh(i\gamma-i\sigma-\sigma_j)}
{\sinh(\sigma_j)\sinh(i\gamma+\sigma_j)}\right)^{L-2tn}e^{-2ipt}
=\nonumber\\
&&-\prod_{l=1}^n\frac{\sinh(\sigma_j-\sigma_l+i\gamma)}
{\sinh(\sigma_j-\sigma_l-i\gamma)},\;\;\;j=1,\ldots,n.
\end{eqnarray}

These equations are quite distinct from the corresponding spectral parameter 
equations 
derived for the row-to-row transfer matrix of the six-vertex model 
\cite{lieb,baxter}. 
Since no numerical 
analysis of the roots for this type of equations were reported in the literature,
we 
 made an extensive numerical study of  them for finite 
values of $L$ and several values of the anisotropy $\Delta$. 
In the particular case where $\Delta=0$ ($\gamma=\frac{\pi}{2})$ 
these equations can be solved analytically. Solutions of these equations 
are obtained by the Newton method by using the distribution of roots 
$\{\sigma_i\}$ at $\Delta=0$ as the starting point to obtain the 
corresponding roots at other values of $\Delta \neq 0$. 
Our numerical analysis shows that the eigenspectrum of $T_{D-D}$ is 
formed by real or complex-conjugated pairs of roots ensuring that the 
partition function ($\ref{eq:e5})$ is a real number. We verified that 
the eigenvalue with highest real part,  belonging to  the sector with $n$ arrows,  
is real and corresponds to a zero momentum eigenstate $(p=0)$. 
The distribution of roots $\{\sigma\}$ corresponding to this eigenvalue 
has a fixed imaginary part, that depends on $\sigma$ and $\gamma$, 
and a symmetrically distributed real part, i. e, 

\begin{eqnarray}
\label{eq:r8}
&&\mbox{Im}(\sigma_j)=\frac{\gamma-\sigma}{2},\;\;\mbox{Real}
(\sigma_j)=\mbox{Real}(\sigma_{n-j}),\nonumber\\
&&j=1,\ldots,n.
\end{eqnarray}
We have also verified, for all sectors, the occurrence of several other 
real eigenvalues. In these cases the corresponding roots $\{\sigma_i\}$ 
have imaginary parts given either by
$\frac{\gamma-\sigma}{2}$ or $\frac{\gamma-\sigma}{2}-\frac{\pi}{2}$. 

In the next section we are going to explore the topology of the above 
distribution of roots in order to extract the bulk limit $(L\rightarrow \infty)$ 
of the interacting five-vertex model.

\section{Thermodynamic Limit}

We  shall obtain in this section the free energy 
$f(\gamma,\sigma,\delta)$ of the interacting five-vertex model in the bulk 
limit $L\rightarrow \infty$. The spectral parameter equations 
($\ref{eq:e43})$ or ($\ref{eq:r7})$ tell us that this interacting 
model with parameter $t$ ($t=0,1,2,\ldots$), on a lattice of size $L$ 
and density of arrows $\rho=\frac{n}{L}$,  is equivalent to a 
six-vertex model ($t=0$) on an effective lattice size $L'=L-2nt$ 
and density $\rho'=\frac{n}{L'}=\frac{n}{L-2tn}$. 
This means that in principle we can derive  the thermodynamic 
properties of the interacting five-vertex models by considering 
the many  sectors, with fixed density of arrows, of the six-vertex ($t=0$)
model defined in the helical geometry of Fig.~\ref{fig1}b.

Since the  relationship among the interacting 
five-vertex models with distinct values of the 
parameters $t$ occurs on the eigensectors with distinct numbers $n$ of 
arrows, we need  to evaluate the highest 
eigenvalue of $T_{D-D}$ on the eigensectors with arbitrary density 
of arrows $\rho = \frac{n}{L}$. The free energy per site in the bulk 
limit $f(\gamma,\sigma,\delta)$ will be obtained from the largest 
eigenvalue of $T_{D-D}$. Due to  ($\ref{eq:r2}$) the eigensector 
containing this eigenvalue will  have a density of arrows  
$\rho = \frac{n}{L}$ that depends on the particular value of 
the parameter $\delta$, i. e,

\begin{eqnarray}
\label{eq:t1}
f(\gamma,\sigma,\delta)=\mbox{min}_{\rho}f_{\rho}(\gamma,\sigma,\delta),
\end{eqnarray}
 where

\begin{eqnarray}
\label{eq:t2}
f_{\rho}=\lim_{n\rightarrow \infty,L\rightarrow \infty,\rho=\frac{n}{L}}
-k_BT\left(\sum_{j=1}^n\frac{\mbox{ln}\lambda_j}{L}+\rho\mbox{ln}\delta
\right),
\end{eqnarray}
 and $\{\lambda_j\}$ are the roots of the equations ($\ref{eq:r3}$) 
 corresponding to the largest eigenvalue of $T_{D-D}$ in the sector 
 with $n=\rho L$ arrows.

The numerical analysis of previous sections  (see ($\ref{eq:r6}$) and 
($\ref{eq:r8}$)) indicates that the roots  $\{\lambda_j\}$ in ($\ref{eq:t2}$) 
correspond to a state with zero momentum and are given by

\begin{eqnarray}
\label{eq:t3}
&&\lambda_j=\frac{\sinh(i\gamma-\sigma_j)}{\sinh\sigma_j},\nonumber\\
&&\sigma_j=v_j+i\frac{\gamma-\sigma}{2},\;\;\; v_j=v_{n-j},j=1,\ldots,n,
\end{eqnarray}
 where $v_j$ are real numbers. Inserting the roots ($\ref{eq:t3}$) in 
($\ref{eq:r7}$) we verify that both 
sides of this last equation are unimodular. Using the relations

\begin{eqnarray}
\frac{\sinh(z-iw)}{\sinh(z+iw)}=e^{-i\theta(z,w)},
\end{eqnarray}

\begin{eqnarray}
\label{eq:t4}
\theta(z,w)=\arctan\left(\frac{2\tanh z\cot w}{1-\tanh^2 z\cot^2 w}
\right),
\end{eqnarray}
 we can replace ($\ref{eq:r7}$) by the equations

\begin{eqnarray}
\label{eq:t5}
&&(L-2nt)(\theta_1(v_j)+\theta_2(v_j))=\nonumber\\
&&2\pi I_j+\sum_{l=1}^n\theta_3(v_l-v_j),\;\;\; j=1,\ldots ,n,
\end{eqnarray}
 where

\begin{eqnarray}
\label{eq:t6}
&&\theta_1(v)=\theta(v,\frac{\gamma+\sigma}{2}),\;\;\; \theta_2(v)=\theta(v,
\frac{\gamma-\sigma}{2}),\nonumber\\
&&\theta_3(v)=\theta(v,\gamma),
\end{eqnarray} 
and $I_j$ are integers or half-odd integers depending 
 if $n$ is odd or even, respectively. 
 Our numerical analysis on small lattices  indicates that those integers are 
given by

\begin{eqnarray}
\label{eq:t7}
I_j=j-\frac{n+1}{2},\;\;\;j=1,\ldots,n.
\end{eqnarray}

In deriving the equations ($\ref{eq:r7}$) and ($\ref{eq:t5}$) it is better 
to consider separately the cases where  the number $n_h = 2L-n(2t+1)$ of 
empty places excluded 
by the hard-core interactions of the arrows (holes)  satisfies 
$n_h \geq n $ and $n_h <n \leq \frac{2L}{2t+1}$, or equivalently,
 $0\leq n \leq \frac{L}{t+1}$ or $\frac{L}{t+1}<n\leq
\frac{2L}{2t+1}$.
If $\frac{L}{2t+1}<n\leq \frac{2L}{2t+1}$, we start again our matrix product 
ansatz ($\ref{eq:e9}$) by associating  now the matrices 
$A^{(1)}$ and $A^{(2)}$ not to the arrows, as before, but to the 
empty places excluded by the hard-core sizes of the arrows (holes). 
The spectral parameter equations we obtain are the same as 
($\ref{eq:r7}$) and ($\ref{eq:t5}$), except that the number $n$ of 
variables $\{\sigma_j\}$ and $\{v_j\}$,  on the right-hand side of 
these equations, is replaced by $ n_h =2L-n(2t+1)$.

In order to calculate the free energy ($\ref{eq:t2}$), as usual, we assume 
that as $L\rightarrow \infty$ the roots $\{v_j\}$ tend to an uniform 
distribution 
on a symmetrical interval of the real axis, $-Q\leq v \leq Q$. 
Let us denoted by $2(L-2nt)R(Q|v)dv$ the number of roots $\{v_j\}$ on the 
interval 
$dv$, in the sector with density of roots $\frac{n}{L}$, i. e,

\begin{eqnarray}
\label{eq:t8}
&&\int_{-Q}^{Q}R(Q|v)dv=\nonumber\\
&&\cases{
\frac{n}{2(L-n(s-1))}=\frac{\rho}{2(1-\rho t)}, 
&  $0\leq\rho\leq \frac{1}{(t+1)}$\,, \cr
\frac{2L-(2s-1)n}{2(L-n(s-1))}=\frac{2-(2t+1)\rho}{2(1-\rho t)}, 
&$\frac{1}{(t+1)}<\rho\leq\frac{2}{(2t+1)}.$\, \cr}
\end{eqnarray}
In terms of this density of roots the function $f_{\rho}$ in 
($\ref{eq:t2}$) is given by

\begin{eqnarray}
\label{eq:t9}
&&f_{\rho}=-2k_BT\rho \ln \delta -2k_BT(1-2\rho t)\nonumber\\
&&\times \int_{-Q}^{Q}R(Q|v)\ln  \left| 
\frac{\sinh(i\frac{(\gamma+\sigma)}{2}-v)}
{\sinh(i\frac{(\gamma-\sigma)}{2}+v)}\right|dv .
\end{eqnarray}
In order to obtain $R(Q|v)$ we notice from ($\ref{eq:t5}$) 
and ($\ref{eq:t7}$) 
that for a given value $v_j$ of the variable $v_j$ the integer 
$I_j+\frac{n+1}{2}$ gives the number of variables $v_l$ with $l<j$. 
 Using this fact ($\ref{eq:t5}$) can be written as \cite{baxter} 

\begin{eqnarray}
\label{eq:t10}
&&(L-2tn)(\theta_1(v) + \theta_2(v)) =  
-\pi(n+1)\nonumber\\
&&+4\pi(L-2tn)\int_{-Q}^vR(Q|v')dv'\nonumber\\
&&+2(L-2tn)\int_{-Q}^Q\theta_3(v^{'}-v)R(Q|v')dv'.
\end{eqnarray}
Taking the derivative of this last expression, with respect to 
 $v$, we obtain  equations that are conveniently expressed 
in the trigonometric regime, $-1\leq \Delta=-\cos \gamma \leq 1$, by

\begin{eqnarray}
\label{eq:t11}
&&R(Q|v)=\nonumber\\
&&\frac{1}{2\pi}\frac{\sin(\sigma+\gamma)}{\cosh(2v)-
\cos(\sigma+\gamma)}
+\frac{1}{2\pi}\frac{\sin(\gamma - \sigma)}{\cosh(2v)-
\cos(\gamma-\sigma)}\nonumber\\
&&-\frac{1}{\pi}\int_{-Q}^Q\frac{\sin(2\gamma)}{\cosh(2(v-v'))-
\cos(2\gamma)}R(Q|v')dv'.
\end{eqnarray}
In the hyperbolic regime, where $\Delta <-1$, it is better to 
change $\gamma \rightarrow -i\lambda$, $\sigma \rightarrow -i\sigma$, 
$v \rightarrow -iv$, and the equations are

\begin{eqnarray}
\label{eq:t12}
&&R(Q|v)=\nonumber\\
&&\frac{1}{2\pi}\frac{\sinh(\sigma+\lambda)}{\cosh(\sigma+\lambda)-
\cos(2v)}
+\frac{1}{2\pi}\frac{\sinh(\lambda-\sigma)}{\cosh(\lambda-\sigma)-
\cos(2v)}\nonumber\\
&&-\frac{1}{\pi}\int_{-Q}^Q\frac{\sinh(2\lambda)}{\cosh(2\lambda)-
\cos(2(v-v'))}R(Q|v')dv',\nonumber\\
&&\Delta=-\cosh \lambda <-1.
\end{eqnarray}

The coupled integral equations ($\ref{eq:t8}$) and ($\ref{eq:t11}$) 
or ($\ref{eq:t8}$) and ($\ref{eq:t12}$) give us, for  a given density 
of roots $\rho$, the parameter $Q=Q(\rho)$ and the density of roots 
$R(Q|v)$. Solving numerically ($\ref{eq:t8}$) and ($\ref{eq:t11}$) 
or ($\ref{eq:t8}$) and ($\ref{eq:t12}$) and inserting the result in 
($\ref{eq:t9}$),  we obtain $f_{\rho}=f_{\rho}(\gamma,\sigma,\delta)$. 
It is important to stress (see ($\ref{eq:t1}$)) that for  given 
values of the parameters 
$\gamma$, $\sigma$ and $\delta $ the free energy is given by the 
minimum value of $f_{\rho}$. In Fig.~\ref{fig7}  we show, for 
$\gamma=\frac{\pi}{3}$ 
$(\Delta=-\frac{1}{2})$, $\sigma=0.1$, $\delta=1$ the curves $f_{\rho}$ 
for some values of the parameters $t$. As we see on this figure 
the minimum value of $f_{\rho}$ that gives the free energy occurs, 
for fixed values of the parameters $\gamma$, 
$\sigma$ and $\delta$, on sectors whose density of arrows $\rho$ depends 
 on the parameters $t$ ($t=0,1,2,\ldots)$. In the case where $t=0$ 
the choice $\delta=1$ gives us the symmetric six-vertex model on the 
helical geometry (Fig.~\ref{fig1}b). The largest eigenvalue 
of $T_{D-D}$ occurs, in this case,  in the sector with density $\rho=1$. 
In Fig.~\ref{fig8} we show, for some values of $t$, the free energy 
$f=\mbox{min}_{\rho}f_{\rho}$ as a function of the parameter $\delta$.

\begin{figure}[h!]
\centering
{\includegraphics[angle=0,scale=0.335]{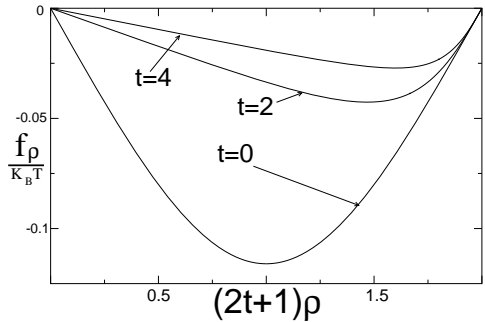}}
\caption{
Functions $f_{\rho}/K_B T$, defined on (\ref{eq:t2}), as a function of $\rho$, 
 for the interacting vertex model with $t=0,2$, and $4$. These curves are  
evaluated for 
  $\Delta = - \frac{1}{2}$, $\sigma =0.1$ and 
$\delta=1$. In this figure the values on the horizontal axis are 
multiplied by $(2t+1)$ to represent the curves on the same scale. The 
minimum values of these functions give the free energy.}
\label{fig7} 
\end{figure}

\begin{figure}[h!]
\centering
{\includegraphics[angle=0,scale=0.4033 ]{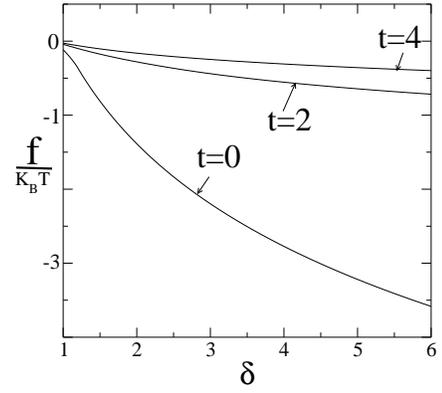}}
\caption{
Free energy, in units of $K_B T$, as a function of $\delta$, for $\delta \geq 
1$ for the 
interacting vertex models with $t=0,2$, and $4$. The curves shown are for the 
anisotropy  
 $\Delta = - \frac{1}{2}$ and parameter $\sigma=0.1$.}
\label{fig8} 
\end{figure}

As we have already mentioned the spectral parameter equations 
($\ref{eq:r3}$) were not analyzed even for the standard 
symmetric six-vertex model. This case corresponds to the case 
where $\delta = 1$ and $t=0$ ($s=1$). Our numerical results (see e.g. 
Fig.~\ref{fig7}) 
indicate 
that in this case the parameter $Q$ in 
($\ref{eq:t8}$)-($\ref{eq:t12}$) 
takes the value $Q\rightarrow \infty$ and $Q=\pi$ for the 
trigonometric ($-1\leq\Delta\leq 1)$ and the hyperbolic 
($\Delta < -1)$ regimes, respectively. We can use in this case Fourier 
transforms in order to solve the coupled integral equations 
($\ref{eq:t8}$) and ($\ref{eq:t11}$) or ($\ref{eq:t8}$) 
and ($\ref{eq:t12}$). After some straightforward algebra we 
obtain for $-1\leq \Delta=-\cos \gamma <1$, $-\gamma <\sigma < \gamma$,

\begin{eqnarray}
\label{eq:t13}
R(Q|v)=R(\infty|v)=\frac{\pi}{4\gamma}
\frac{\cosh(\frac{v\pi}{\gamma})\cos(\frac{\pi\sigma}{2\gamma})}
{\cosh(\frac{2v\pi}{\gamma})+\cos(\frac{\pi\sigma}{\gamma})},
\end{eqnarray}
that gives   the free energy

\begin{eqnarray}
\label{eq:t14}
&&f=
-k_BT\int_0^{\infty}\frac{2}{\gamma}\frac{\cosh(\frac{v\pi}{\gamma})
\cos(\frac{\pi\sigma}{2\gamma})}{\cosh(\frac{2v\pi}{\gamma})+
\cos(\frac{\pi\sigma}{\gamma})}\nonumber\\
&&\times\ln\left(
\frac{\cosh(2v)-\cos(\gamma+\sigma)}{\cosh(2v)-
\cos(\gamma-\sigma)}\right)dv.
\end{eqnarray}
On the other hand, for  $\Delta=-\cosh \lambda < -1$, $\lambda >0$, $-\lambda < \sigma < 
\lambda$, 

\begin{eqnarray}
\label{eq:t15}
&&R(Q|v)=R(\pi|v)=\nonumber\\
&&\frac{1}{4\pi}\int_{-\pi}^{\pi}\cosh(\frac{\sigma x}{2}){\mbox{sech}}
(\frac{\lambda x}{2})e^{-ixv}dx
\nonumber 
\end{eqnarray} 
 and
\begin{eqnarray}
\label{eq:t16}
f=-k_BT(\sigma+\sum_{m=1}^{\infty}\frac{(-1)^m}{m}
\sinh(2m\sigma){\mbox{sech}}(m\lambda)e^{-m\lambda}). 
\nonumber 
\end{eqnarray}

In table \ref{table1} we present on the last line the numerical evaluation 
of the integral $(\ref{eq:t14})$ for some values of 
$\Delta=-\cos \gamma <1$. 
In order to compare with the results for finite values of $L$ we 
also show on this table the results obtained by solving directly 
the spectral parameter equations $(\ref{eq:r7})$. 
The free energy per site in the bulk limit should be  independ of  the 
geometry of the lattice. Indeed the results of table \ref{table1}, 
obtained for the symmetric  six-vertex model in the helical 
geometry,  coincide with those  obtained from the row-to-row 
transfer matrix in the usual geometry (Fig.~\ref{fig1}a). 
It is however interesting to observe that although the final 
results are the same,  the integrand in (\ref{eq:t14}) for the 
helical case (Fig.~\ref{fig1}b) is distinct from the corresponding one in 
the geometry of Fig.~\ref{fig1}a (see Eq. (8.8.19) in \cite{baxter})
\normalsize




\begin{table}[hbt]
\centering
\begin{tabular}{|c|c|c|c|c|} 
 \hline
$L$       &$\gamma=\frac{2\pi}{3}$       &$\gamma=\frac{\pi}{2}$   
&$\gamma=\frac{\pi}{3}$   \\ \hline
\hline
6       &\,0.03512818236       &\,0.06453944019  &\,0.11709717612     \\ 
10      &\,0.03477636310       &\,0.06406728711  &\,0.11638357858     \\ 
18      &\,0.03464007645       &\,0.06388499114  &\,0.11610862696     \\ 
34      &\,0.03459635046       &\,0.06382657413  &\,0.11602059391    \\ 
66      &\,0.03458384714       &\,0.06380987622  &\,0.11599543821    \\ 
130     &\,0.03458049496       &\,0.06380539993  &\,0.11598869526    \\ 
$\infty$ &\,0.03457933094     &\,0.06380384561 &\,0.11598635395      \\ \hline
\end{tabular}
\caption{
Free energy of the six-vertex model ($t=0$) on the distorted lattice of 
Fig.~\ref{fig1}b for
some values of the anisotropy $\Delta = -\cos \gamma$ and lattice sizes $L$. 
The values 
are obtained by solving (\ref{eq:r7}) with $\sigma =0.1$ and using 
(\ref{eq:r6}) and 
(\ref{eq:t2}) with $\delta=1$. The last line gives the bulk limit obtained 
by solving (\ref{eq:t14}).}
\label{table1}
\end{table}

\section{Conformal invariance and operator content}

In this section, by exploring the conformal invariance of the 
infinite system, we shall  obtain the critical behavior of 
the massless phases of the interacting five-vertex models 
introduced on this paper. The results of the last section tell us 
that for $\Delta < 1$ the eigensector of $T_{D-D}$ presenting  the 
largest eigenvalue,  changes continuously with the value of the 
interacting parameter $\delta$. This imply that for $\Delta < 1$ the 
introduced interacting models with arbitrary values of 
$t=0,1,2,\ldots$ are on a critical phase, as we change 
continuosly the parameter $\delta$ in the bulk limit ($L\rightarrow \infty)$.
 This behavior for the case $t=0$ ($s=1$) is expected since, 
as we discussed in earlier sections, the  interacting 
model recovers the six-vertex model.    In this case the interacting parameter 
$\delta$ plays the role of a chemical potential (or electric field) 
controlling the density of arrows. The critical behavior of this 
six-vertex model follows from that of the XXZ chain in the presence 
of an external magnetic field \cite{magfield}. The region where $\Delta > 1$ 
(ferroelectric region) is non-critical for any value of $t$ and $\delta$. 
The largest eigenvalue of $T_{D-D}$ belongs to the sector with no arrows
 $(\rho=0)$, for $\delta < 1$, or to the sector with the largest possible 
density of arrows $(\rho=\frac{2}{2t+1})$ for $\delta \geq 1$.

The conformal anomaly and anomalous dimensions (related to the 
critical exponents) of the underlying conformal field theory 
governing the critical fluctuations on the critical phase of 
our models are going to be calculated from the finite-size corrections
 of the eigenvalues of the transfer matrix $T_{D-D}$, 
or the related Hamiltonian $H=-\ln (T_{D-D})$. The leading 
behavior, as $L\rightarrow \infty$, of the smallest eigenvalue 
$F_0(L)$ of $H$ (ground-state energy) \cite{blote}:

\begin{eqnarray}
\label{eq:c1}
F_0(L)=F_0(\infty)-\frac{\pi v_sc}{6L}+o(L^{-1})
\end{eqnarray}
 gives  the conformal anomaly $c$ and the sound velocity $v_s$. 
Moreover, to each primary operator $\phi$ with dimensions 
$x_{\phi}$, in the operator algebra of the critical system, 
there exists an infinite tower of eigenstates of 
$H=-\mbox{ln}(T_{D-D})$, whose eigenvalues $F^{\phi}_{m,m^{'}}(L)$ 
behaves asymptotically as \cite{cardy} 

\begin{eqnarray}
\label{eq:c2}
F^{\phi}_{m,m'}(L)=F_0(L)+\frac{2\pi v_s}{L}(x_{\phi}+m+m')+o(L^{-1}),
\end{eqnarray}

\noindent with $m,m'=0,1,2\ldots$.

Before considering our model with general values of 
$t$ $(t=0,1,2\ldots)$ and $\delta$, let us consider  the 
special case where $t=0$ and $\delta=1$. In this case we recover the standard 
symmetric six-vertex model in the absence of electric fields, 
but for  the helical geometry of Fig.~\ref{fig1}b.
Similarly, as in the standard geometry of Fig.~\ref{fig1}a \cite{bogoliubov}, 
our numerical and analytical analysis indicate that the model 
is critical for $-1<\Delta <1$, and massive for  $|\Delta|>1$. 
In agreement with earlier studies of the six-vertex on the 
standard geometry, and of  the XXZ quantum chain \cite{ab2},  the critical 
fluctuations  
in the region $-1< \Delta =-\cos \gamma <1$,  are governed by a 
conformal field theory with central charge $c=1$ and anomalous 
dimensions given by

\begin{eqnarray}
\label{eq:c3}
x_{n,m}=n^2x_p+\frac{m^2}{4x_p},\;\;\; n,m=0,\pm 1,\pm 2,
\end{eqnarray}
 where 

\begin{eqnarray}
\label{eq:c4}
x_p=\frac{\pi-\gamma}{2\pi}.
\end{eqnarray}

Our numerical and analytical analyses show  that distinctly from 
the row-to-row transfer matrices in the standard geometry where 
the sound velocity is  unity \cite{cardy,batchelor,fn4},  
 in the helical 
geometry 
one has $v_s=v_s(\gamma,\sigma)\neq 1$. The sound velocity 
from ($\ref{eq:c1}$) can be calculated from the bulk limit of 
the finite-size sequence

\begin{eqnarray}
\label{eq:c5}
v^{(L)}_s=[F_0(L)-F(\infty)]\frac{6L}{\pi}.
\end{eqnarray}
For $\Delta=0$ $(\gamma=\frac{\pi}{2})$ we can compute 
analytically these finite-size  corrections and obtain

\begin{eqnarray}
\label{eq:c6}
v_s=\frac{1}{2}\tan \frac{\sigma}{2}.
\end{eqnarray}
For $\Delta \neq 0$ our  calculations were done numerically and we 
do not have an  analytical form for $v_s$. In table \ref{table2} we show 
some of the estimates $v_s(L)$ given by $(\ref{eq:c5})$ obtained 
for some values of $\Delta$ at $\sigma=0.1$.
In order to illustrate our numerical evaluation of the dimensions 
(\ref{eq:c3}) we show in table \ref{table3}, for $\sigma=0.1$ and some values 
of $\gamma$, the finite-size estimates for $x_p$, together with the 
expected results (\ref{eq:c4}). These estimates are obtained 
by using in (\ref{eq:c2}) the largest eigenvalues of $T_{D-D}$ 
in the sectors with $n=L$ and $n=L-1$ arrows.

\begin{table}[hbt]
\centering
\begin{tabular}{|c|c|c|c|c|}
 \hline
$L$       &$\gamma=\frac{2\pi}{3}$       &$\gamma=\frac{\pi}{2}$  
&$\gamma=\frac{\pi}{3}$     \\ \hline
\hline
6       &\,0.03773624455       &\,0.05057575759  &\,0.07637450614     \\ 
10      &\,0.03763037006       &\,0.05031361905  &\,0.07586431519     \\ 
18      &\,0.03758898261       &\,0.05021239628  &\,0.07566185294     \\ 
34      &\,0.03757566153       &\,0.05017995742  &\,0.07559488486     \\ 
66      &\,0.03757184873       &\,0.05017068491  &\,0.07557509072    \\ 
130     &\,0.03757082640       &\,0.05016819918  &\,0.07556960190   \\
 \hline
\end{tabular}
\caption{
Finite-size estimators $v_s(L)$ for the sound velocity of the six-vertex model 
($t=0$) on the distorted lattice of Fig.~\ref{fig1}b. The values are given 
for some values of the anisotropy $\Delta = -\cos \gamma$ and lattice sizes 
$L$.}
\label{table2}
\end{table}

\begin{table}[hbt]
\centering
\begin{tabular}{|c|c|c|c|}
\hline
$L$       &$\gamma=\frac{2\pi}{3}$       &$\gamma=\frac{\pi}{2}$  
&$\gamma=\frac{\pi}{3}$   \\  \hline
\hline
6       &\,0.16676687232       &\,0.24712196038  &\,0.32156619180     \\ 
10      &\,0.16669813477       &\,0.24895867881  &\,0.32848712047    \\ 
18      &\,0.16667118823       &\,0.24967533316  &\,0.33163455071    \\ 
34      &\,0.16666377768       &\,0.24990588089  &\,0.33279409868    \\ 
66      &\,0.16666111003       &\,0.24997186077  &\,0.33308163424     \\ 
130     &\,0.16666077565       &\,0.24998955436  &\,0.33328029440     \\ 
$x_p$   &\,0.1666..  &\,0.25  &\,0.3333..    \\ \hline
\end{tabular}
\caption{
Finite-size estimators for the critical exponent $x_p$ of the six-vertex model 
on the distorted lattice ($t=0$) of Fig.~\ref{fig1}b. The values are given for 
some values of the anisotropy $\Delta = - \cos \gamma$ and lattice sizes $L$. 
These estimators are obtained by solving (\ref{eq:r7}).}
\label{table3}
\end{table}

Let us consider the general case where $t=0,1,2\ldots$ and $\delta>0$. 
In this case we have a massless phase for $\Delta <1$. 
For a given value of $\delta$ the density of arrows $\rho$, 
in the bulk limit, is obtained by the condition 
$\frac{df_{\rho}}{d\rho}=0$, where $f_{\rho}$  is 
given by $(\ref{eq:t9})$. As we change $\delta$, from zero 
to infinite,  $\rho$ increases from $0$ to $\frac{2}{2t+1}$. 
Our numerical analysis indicates that the underlying conformal field 
theory,  as in the case $t=0$ and $\delta=1$, has also a central charge $c=1$ 
and the Coulomb gas type of conformal dimensions given in 
$(\ref{eq:c3})$, but with a value of $x_p$ that depends 
on $\delta$,$\gamma$ and $\sigma$. The parameter $x_p$, that can be estimated 
 from the 
finite-size corrections of the smallest eigenvalues 
of $H=-\mbox{ln}(T_{D-D})$, can be calculated analytically. This is done by  
applying to our relations $(\ref{eq:t8})$-$(\ref{eq:t12})$ 
of last section,  the method   used in \cite{woyna} for the XXZ 
in a magnetic field. We obtain

\begin{eqnarray}
\label{eq:f2}
x_p=\frac{1}{4}(1-t\rho)^{-2}\eta^{-2}(Q|v)
\end{eqnarray}
 where, for $-1\leq\Delta=-\cos \gamma \leq1$,

\begin{eqnarray}
\label{eq:f3}
\eta(Q|v)=1-\frac{1}{\pi}\int_{-Q}^Q\frac{\sin(2\gamma)\eta(Q|v')}
{\cosh(2(v-v')-\cos(2\gamma)}dv',
\end{eqnarray}
 and, for $\Delta=-\cosh \lambda <-1$,

\begin{eqnarray}
\label{eq:f4}
\eta(Q|v)=1-\frac{1}{\pi}\int_{-Q}^Q\frac{\sinh(2\lambda)\eta(Q|v')}
{\cosh(2\lambda)-\cos(2(v-v')}dv'.
\end{eqnarray}
 For a given density of arrows $\rho$ the parameter
$Q=Q(\rho)$ is obtained by solving the coupled integral equations
$(\ref{eq:t8})$ and $(\ref{eq:t11})$ for $-1<\Delta < 1$, and
$(\ref{eq:t8})$ and $(\ref{eq:t12})$ for $\Delta=-\cosh \lambda <-1$.
 Similar integral equations happen for the XXZ chain with hard-core
exclusion effects \cite{mac}.

At $\Delta =0$ Eq. (\ref{eq:f3}) gives $\eta(Q|v) = 1$, and from (\ref{eq:f4}) 
we obtain the
curve
\begin{equation}
\label{eqfree}
x_p = \frac{1}{4(1-t\rho)^2}, \;\;\;\ \Delta =0.
\end{equation}

For fixed values of $t,\sigma$ and $\gamma$ (or $\lambda$) the
interacting parameter $\delta$ will fix the density of arrows $\rho$
as well as  the parameters $Q=Q(\rho)$ appearing in
$(\ref{eq:t8})$, $(\ref{eq:t11})$
(or $(\ref{eq:t12})$)  and $(\ref{eq:f2})$-$(\ref{eq:f4})$.
In Fig.~\ref{fig9}   we show the curves of $x_p$ as a function of $\rho$ 
at $\Delta = -\frac{1}{2}$ and
for some fixed values of $t$. 
In Fig.~\ref{fig10} and Fig.~\ref{fig11} the values of  $x_p$
as a function of $\rho$ are 
shown for the interacting model with parameter $t=2$. The curves of 
Fig.~\ref{fig10} and 
Fig.~\ref{fig11} are for the regimes $-1 \leq \Delta \leq 1$ and $\Delta <-1$, 
respectively.

As we see in Fig.~\ref{fig10} and Fig.~\ref{fig11} the curves have a  common
intercept $x_p=1/4$ at $\rho=0$ and $x_p=25/4=6.25$ at $\rho =2/5 = 0.4$.
In fact our numerical results, for arbitrary values of $t$ and $\Delta$, shows 
that
at low density ($\rho \rightarrow 0$) and high density of arrows ($\rho 
\rightarrow
\frac{2}{2t+1}$) the anomalous dimension $x_p$ take  the values $x_p=1/4$
and $x_p = \frac{(2t+1)^2}{4}$, respectively. This happens since at those 
extreme limits
either the density of arrows or the density of holes is low and the 
anisotropy
$\Delta$, that is only important for the arrows (holes) at the closest 
positions,
plays no role.

We also see in Fig.~\ref{fig11} the appearance of a peak ($x_p = 6^2/4 =9$) at 
the  density
$\rho =  \frac{1}{3}$, that is independent of the anisotropy $\Delta$.
For general values of $t$ and $\Delta <-1$ this peak occurs at the
"half-filled" density $\rho=\frac{1}{t+1}$, for any $\Delta <-1$. 
 This happens because, for $\Delta <-1$, the system at this "half-filling" has a 
mass gap. 
An infinitesimal
change of $\rho$ (induced by a change of $\delta$) destroys this mass gap. 
Since
around this density the main effect is the disappearance of the gap, the 
critical exponent
 $x_p$ does not depend on the $\Delta$ value.
 A similar effect is also expected in the
region $\Delta <-1$ of the XXZ or  six-vertex model on the presence of an 
external
magnetic field.

\begin{figure}[h!]
\centering
{\includegraphics[angle=0,scale=0.30]{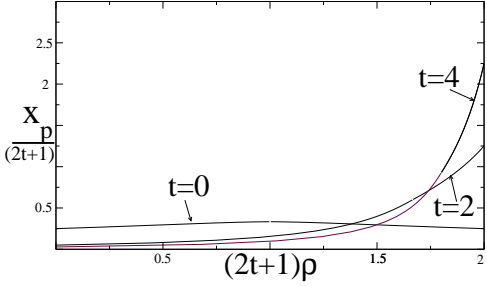}}
\caption{
The critical exponent $x_p$, as a function of $\rho$, for the interacting 
vertex models with $\Delta = -\frac{1}{2}$ and 
interacting parameter $t=0,2$ and $4$. In the figure the horizontal and 
(vertical) axis are 
multiplied (divided) by $2t+1$ in order to represent the curves on the same 
scale.}
\label{fig9} 
\end{figure}

\begin{figure}[h!]
\centering
{\includegraphics[angle=0,scale=0.46]{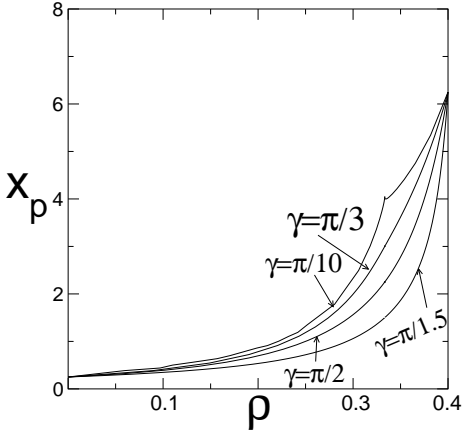}}
\caption{
The critical exponent $x_p$, as a function of $\rho$, for the vertex model with 
interacting parameter $t=2$ and some values of the anisotropy 
$\Delta = - \cos \gamma \leq -1$.}
\label{fig10} 
\end{figure}

\begin{figure}[h!]
\centering
{\includegraphics[angle=0,scale=0.43]{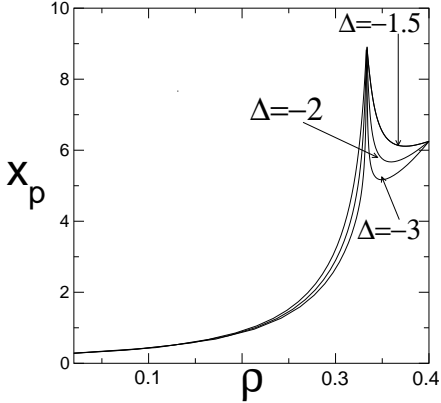}}
\caption{
The critical exponent $x_p$, as a function of $\rho$, for the vertex model 
with interacting parameter $t=2$ and some values of the anisotropy 
$\Delta <-1$.}
\label{fig11} 
\end{figure}

\section{Conclusions}

We have introduced in this paper a special family of exact solvable 
interacting five-vertex models. Besides the usual 
nearest-neighbor 
interactions imposed by the lattice connectivity, such models also present  hard-core 
interactions along one of the diagonals of the lattice. The range of the 
additional interactions depend on a fixed parameter $t$ ($t=1,2,\ldots$). 
These vertex models can also be interpreted as if the arrows entering in  the 
vertex configurations have an effective hard-core size $s = 2t+1$. This 
interpretation allows us to identify the extension of these models 
to the case $t=0$ ($s=1$) as 
the standard six-vertex model. In this sense, although our interacting models 
have only five vertex,  they are generalizations of the standard six-vertex 
model. 

Although  the exact solution of the transfer matrix are usually obtained 
through the 
Bethe ansatz, we have derived the solution of these models by using the matrix 
product ansatz introduced in \cite{alclazo}, to perform   the exact diagonalization of 
quantum 
 chains. This first application indicates that this ansatz is also suitable 
 for transfer matrix calculations.  
 
 Due to the extra interactions along the  diagonals of the lattice,  the 
solution 
 of these models 
 was simplified by considering the diagonal-to-diagonal transfer 
 matrix. A numerical analysis of the spectral parameter equations for small 
 lattices enabled the calculation of  the free energy of the models in the 
 bulk limit. This calculation was done on a symmetric version of the model 
 that recovers for $t=0$ the symmetric six-vertex model. Our results shows 
 that the models, for any  value of $t$ ($t=0,1,2,\ldots$), exhibit a 
 massless phase for arbitrary values of the anisotropy $\Delta$ given by 
 (\ref{eq:r4}). Exploring the consequences of the conformal invariance of 
 the infinite system we also calculate, on the massless phase, the conformal 
 anomaly, the sound velocity and the operator content of the underlying 
 field theory governing the long-distance critical fluctuations.  Underlying these 
 models there is  a $c=1$ conformal field theory whose operators have 
 a Coulomb gas type of anomalous dimensions $x_{n,m}$ (see (\ref{eq:c3})). 
 The compactification ratio defining these dimensions varies continuosly 
 with the free parameters of the model. Our analytical and numerical 
 calculations show that the effect of the interacting parameter $t$ is 
 to increase the critical exponents as $t$ increases, 
 while the extra interaction energy, related to the parameter 
 $\delta$, plays the role of a chemical potential controlling the number of 
 arrows on the ground state eigenfunction.

\section*{Acknowledgments}
We thank V. Rittenberg for helping us to find the representation (\ref{repre}) 
and L. A. Ferreira for fruitfull discussions and a carefull reading of the 
manuscript. This work has been partially supported by the Brazilian agencies FAPESP and 
CNPq.


\begin{thebibliography}{99}
\bibitem {fn1} This notation is 
 chosen in order to compare these interacting five-vertex models with the 
standard six-vertex model.
\bibitem {fn2} This last case only is allowed in the case of the six-vertex 
 model where $t=0$ and $s=1$.
\bibitem {fn3} The phase $e^{-i2pt}$ in  ($\ref{eq:e43}$) could be obtained by 
 considering  a six-vertex model on the geometry of Fig.~\ref{fig1}b, but with 
a seam 
 with distinct vertex fugacities along the vertical direction.
\bibitem {fn4} 
In the XXZ chain with anisotropy $\Delta = -\cos \gamma $ the sound  
velocity is $v_s = \frac{\pi \sin \gamma}{\gamma}$.
\end{thebibliography}
\end{document}